\documentclass{webofc}
\usepackage[varg]{txfonts}   

\newcommand{\Pom}{\mathbb{P}}
\newcommand{\Ode}{\mathbb{O}}
\newcommand{\Reg}{\mathbb{R}}

\newcommand{\bdPt}{\mbox{\boldmath $dP_{t}$}}
\newcommand{\bqta}{\mbox{\boldmath $q_{t,1}$}}
\newcommand{\bqtb}{\mbox{\boldmath $q_{t,2}$}}
\newcommand{\bpta}{\mbox{\boldmath $p_{t,1}$}}
\newcommand{\bptb}{\mbox{\boldmath $p_{t,2}$}}

\begin{document}
\title{Tensor pomeron, vector odderon and diffractive production of meson and baryon pairs
in proton-proton collisions}
%
%

\author{\firstname{Piotr} \lastname{Lebiedowicz}\inst{1}\fnsep\thanks{\email{Piotr.Lebiedowicz@ifj.edu.pl}} \and
        \firstname{Otto} \lastname{Nachtmann}\inst{2}\fnsep
\thanks{\email{O.Nachtmann@thphys.uni-heidelberg.de}}
\and
        \firstname{Antoni} \lastname{Szczurek}\inst{1}\fnsep
\thanks{\email{Antoni.Szczurek@ifj.edu.pl}}
}

\institute{Institute of Nuclear Physics Polish Academy of Sciences,\\
           ul. Radzikowskiego 152, PL 31-342 Krak{\'o}w, Poland
\and
           Institut f\"ur Theoretische Physik, Universit\"at Heidelberg,\\
           Philosophenweg 16, D-69120 Heidelberg, Germany
          }

\abstract{
We review some selected results of the tensor-pomeron and vector-odderon model of soft high-energy proton-proton scattering and central exclusive production of meson and baryon pairs in proton-proton collisions. We discuss the theoretical aspects of this approach and consider the phenomenological implications in a variety of processes at high energies, comparing to existing experimental data. We consider the diffractive dipion and dikaon production including the continuum and the dominant scalar and tensor resonance contributions as well as the photoproduction processes. The theoretical results are compared with existing CDF experimental data 
and predictions for planned or current LHC experiments, ALICE, ATLAS, CMS, LHCb  are presented. 
}
%
\maketitle

\section{Introduction}
\label{intro}

There is a growing experimental and theoretical interest 
in understanding diffractive processes at high energy in proton-(anti)proton collisions.
A particularly interesting are central exclusive production (CEP) processes 
where all centrally produced particles are detected. 
In the CDF \cite{Aaltonen:2015uva} and the CMS \cite{Khachatryan:2017xsi} experiments 
only large rapidity gaps around the centrally produced dimeson system 
are checked but the forward and backward going protons are not detected.
Preliminary results of similar CEP studies were presented by the ALICE \cite{Schicker:2012nn}
and LHCb \cite{McNulty:2016sor} Collaborations at the LHC
and by the STAR \cite{Sikora:2018cyk} Collaboration at RHIC.
Although such results will have diffractive nature, 
further efforts are needed to ensure their exclusivity.
These measurements are important in the context of resonance production, 
in particular, in searches for the gluon bound states (glueballs).
Future experiments at the LHC will be able to detect all particles
produced in CEP, including the forward and backward going protons.
Feasibility studies for the $p p \to p p \pi^+ \pi^-$ process with 
tagging of the scattered protons as carried out 
for the ATLAS and ALFA detectors are shown in \cite{Staszewski:2011bg}.
Similar possibilities exist with the CMS and TOTEM detectors; see, e.g. \cite{Albrow:2014lrm}.

It was known for a long time that the frequently 
used vector-pomeron model has problems from the point of view of field theory.
Taken literally it gives opposite signs for $pp$ and $\bar{p}p$ total cross sections.
A way how to solve these problems was discussed in \cite{Nachtmann:1991ua}
where the pomeron was described as a coherent superposition of exchanges
with spin $2 + 4 + 6 + ...$.
The same idea is realised in the tensor-pomeron model formulated in \cite{Ewerz:2013kda}
where the soft pomeron exchange can effectively be treated as the
exchange of a rank-2 symmetric tensor.
In \cite{Ewerz:2016onn} it was shown that the tensor-pomeron model 
is consistent with the experimental data 
on the helicity structure of proton-proton elastic scattering at
$\sqrt{s} = 200$~GeV and small $|t|$ from the STAR experiment \cite{Adamczyk:2012kn}.
In Ref.~\cite{Lebiedowicz:2013ika} the tensor-pomeron model was applied to 
the diffractive production of several scalar and pseudoscalar mesons 
in the reaction $p p \to p p M$ and it was shown that this model does quite well
in reproducing the data when available.
In \cite{Bolz:2014mya} an extensive study of the photoproduction reaction
$\gamma p \to \pi^{+} \pi^{-} p$ in the framework
of the tensor-pomeron model was presented.
The resonant ($\rho^0 \to \pi^{+}\pi^{-}$) and non-resonant (Drell-S\"oding)
photon-pomeron/reggeon $\pi^{+} \pi^{-}$ production in $pp$ collisions
was studied in \cite{Lebiedowicz:2014bea}.
The central exclusive diffractive production of $\pi^{+} \pi^{-}$ continuum 
together with the dominant scalar $f_{0}(500)$, $f_{0}(980)$, 
and tensor $f_{2}(1270)$ resonances was studied by us in \cite{Lebiedowicz:2016ioh}.
The experimental data on central exclusive $\pi^{+}\pi^{-}$ production
measured at Fermilab \cite{Aaltonen:2015uva}, CERN \cite{Khachatryan:2017xsi}, 
and RHIC \cite{Sikora:2018cyk}
all show visible structures in the $\pi^{+}\pi^{-}$ invariant mass.
As we discussed in Ref.~\cite{Lebiedowicz:2016ioh}
the pattern of these structures has mainly resonant origin and is very sensitive to
the cuts used in a particular experiment
(usually these cuts are different for different experiments).
The $\rho^{0}$ meson production associated with 
a very forward/backward $\pi N$ system
in the $pp \to pp \rho^{0} \pi^{0}$ and $pp \to pn \rho^{0} \pi^{+}$ processes
was discussed in \cite{Lebiedowicz:2016ryp}.
In \cite{Lebiedowicz:2018eui}, the exclusive diffractive production of the $K^{+} K^{-}$
in the continuum and via the dominant scalar $f_{0}(980)$, $f_{0}(1500)$, $f_{0}(1710)$,
and tensor $f_{2}(1270)$,  $f'_{2}(1525)$ resonances, 
as well as the $K^{+} K^{-}$ photoproduction contributions, were discussed in detail.
In \cite{Lebiedowicz:2018sdt} the $pp \to pp p\bar{p}$ reaction was studied.
Also the central exclusive $\pi^+ \pi^-\pi^+ \pi^-$ production 
via the intermediate $\sigma\sigma$ and $\rho^0\rho^0$ states in $pp$ collisions
was studied in \cite{Lebiedowicz:2016zka}.

\section{Formalism}
\label{formalism}

\begin{figure}[ht]
\centering
(a)\includegraphics[width=4.6cm,clip]{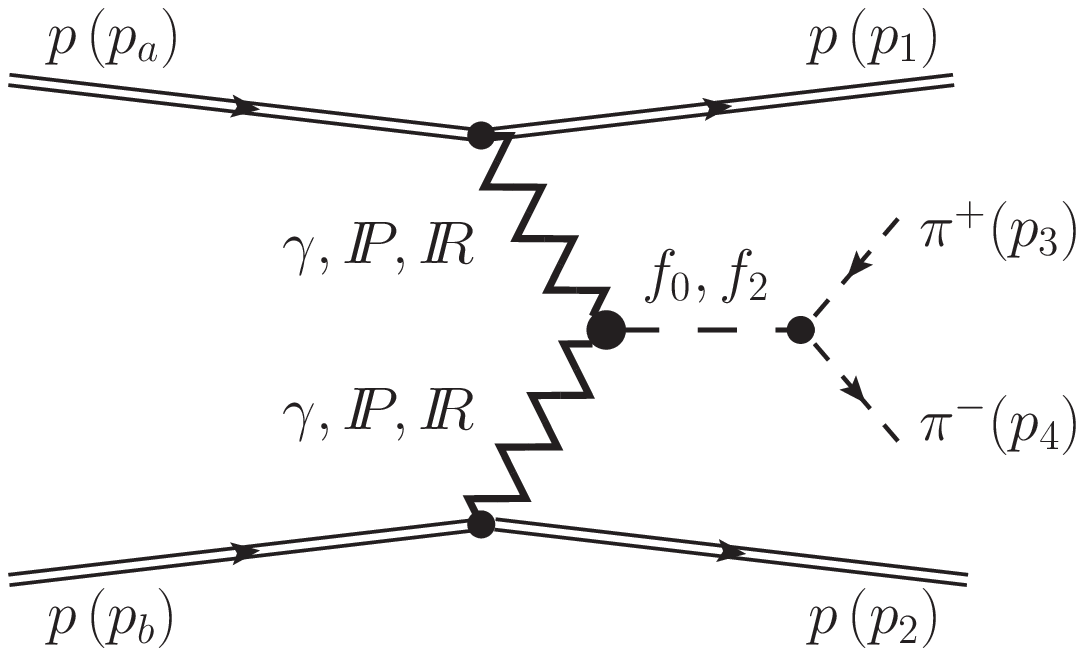}
(b)\includegraphics[width=4.cm,clip]{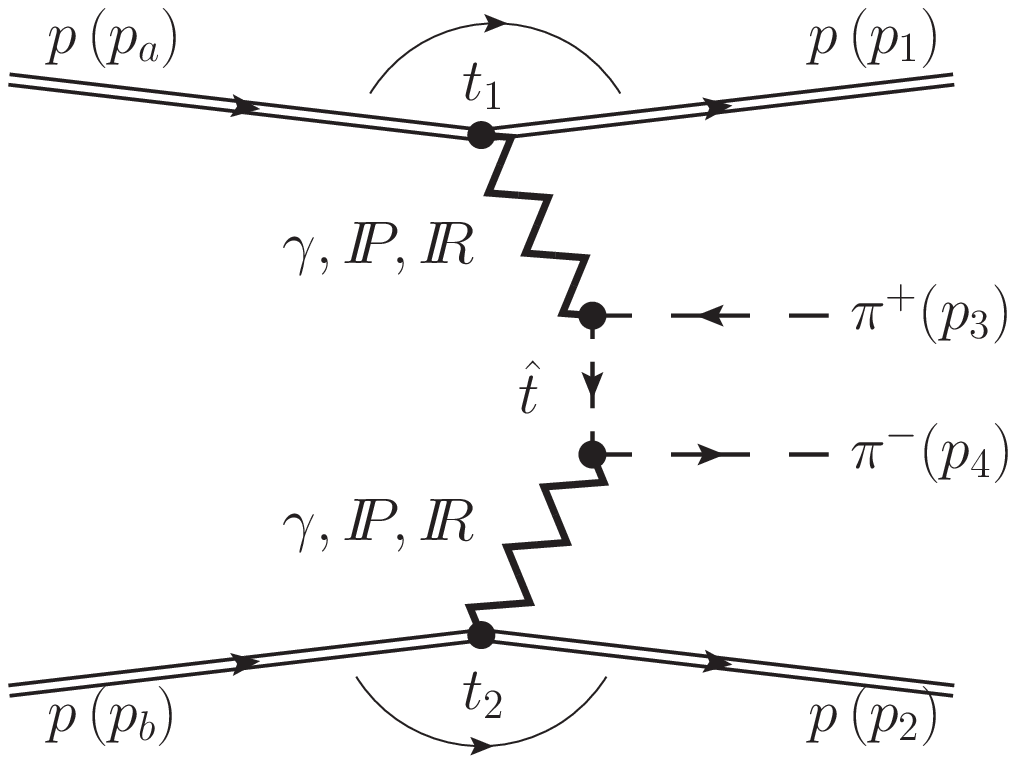}
   \includegraphics[width=4.cm,clip]{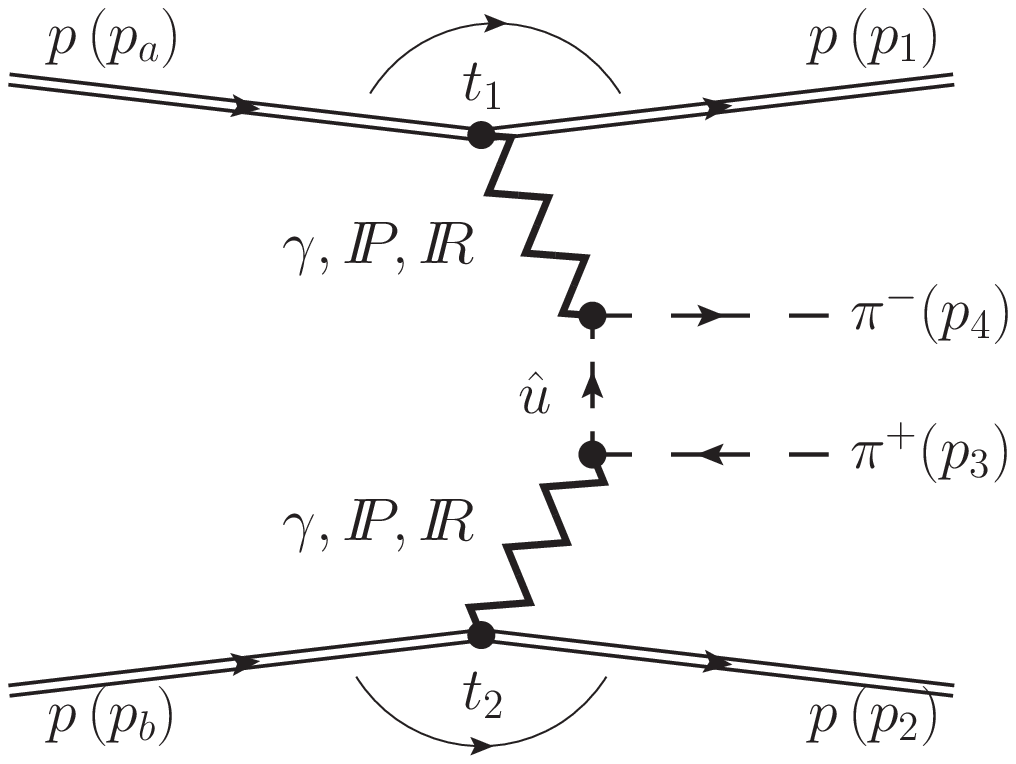}
\caption{
The diagram (a) shows the double-pomeron/reggeon and photon mediated
central exclusive $I^{G}J^{PC} = 0^{+}0^{++}$ and $0^{+}2^{++}$ resonances production
and their subsequent decays into $\pi^+ \pi^-$ in proton-proton collisions.
Two diagrams in panel (b) show exclusive continuum $\pi^+ \pi^-$ production.
}
\label{fig:0}
\end{figure}

We discuss central exclusive production of $\pi^{+}\pi^{-}$ pairs 
in proton-proton collisions at high energies
\begin{eqnarray}
p\,(p_{a}) + p\,(p_{b}) \to p\,(p_{1}) + \pi^{+}\,(p_{3}) + \pi^{-}\,(p_{4}) + p\,(p_{2}) \,,
\label{2to4_reaction}
\end{eqnarray}
where $p_{i}$, indicated in brackets, denote the 4-momenta of the particles.
The generic diagrams for diffractive exclusive $pp \to pp \pi^{+} \pi^{-}$ reaction
are shown in Fig.~\ref{fig:0}.
At high energies the exchange objects to be considered are
the photon $\gamma$, the pomeron $\Pom$, the odderon $\Ode$, 
and the reggeons $\Reg$.
Their charge conjugation and $G$-parity quantum numbers
are listed in Table~I of \cite{Lebiedowicz:2016ioh}.
We treat the $C=+1$ pomeron and the reggeons $\Reg_{+} = f_{2 \Reg}, a_{2 \Reg}$ 
as effective tensor exchanges
while the $C=-1$ odderon and the reggeons 
$\Reg_{-} = \omega_{\Reg}, \rho_{\Reg}$ are treated as effective vector exchanges.
Note that $G$-parity invariance forbids the vertices $a_{2 \Reg} \pi \pi$,
$\omega_{\Reg} \pi \pi$ and $\Ode \pi \pi$.

The total amplitude for the $pp \to pp \pi^{+} \pi^{-}$ reaction
is a coherent sum of continuum amplitudes
and the amplitudes with the $s$-channel 
scalar ($J^{PC} = 0^{++}$) and tensor ($J^{PC} = 2^{++}$) resonances.
The photoproduction contributions, discussed in detail in \cite{Lebiedowicz:2014bea}, 
must be also added coherently at the amplitude level 
and in principle could interfere.

For example, the $\Pom\Pom$-exchange ``Born-level'' amplitude (without absorption effects) 
for $\pi^{+} \pi^{-}$ production through a tensor resonance $f_{2} \to \pi^{+} \pi^{-}$
can be written as
\begin{eqnarray}
&&{\cal M}^{(\Pom \Pom \to f_{2}\to \pi^{+}\pi^{-})}_{\lambda_{a} \lambda_{b} \to \lambda_{1} \lambda_{2} \pi^{+}\pi^{-}} 
=  (-i)\,
\bar{u}(p_{1}, \lambda_{1}) 
i\Gamma^{(\Pom pp)}_{\mu_{1} \nu_{1}}(p_{1},p_{a}) 
u(p_{a}, \lambda_{a})\;
i\Delta^{(\Pom)\, \mu_{1} \nu_{1}, \alpha_{1} \beta_{1}}(s_{1},t_{1}) \nonumber \\
&& \qquad \qquad \qquad \quad \times \;
i\Gamma^{(\Pom \Pom f_{2})}_{\alpha_{1} \beta_{1},\alpha_{2} \beta_{2}, \rho \sigma}(q_{1},q_{2}) \;
i\Delta^{(f_{2})\,\rho \sigma, \alpha \beta}(p_{34})\;
i\Gamma^{(f_{2} \pi\pi)}_{\alpha \beta}(p_{3},p_{4}) \nonumber \\
&& \qquad \qquad \qquad \quad \times \;
i\Delta^{(\Pom)\, \alpha_{2} \beta_{2}, \mu_{2} \nu_{2}}(s_{2},t_{2}) \;
\bar{u}(p_{2}, \lambda_{2}) 
i\Gamma^{(\Pom pp)}_{\mu_{2} \nu_{2}}(p_{2},p_{b}) 
u(p_{b}, \lambda_{b}) \,,
\label{amplitude_f2_pomTpomT}
\end{eqnarray}
where $\lambda_{i} \in \lbrace +1/2, -1/2 \rbrace$
denote the helicities of the nucleons,
$t_{1} = q_{1}^{2} = (p_{1} - p_{a})^{2}$, $t_{2} = q_{2}^{2} = (p_{2} - p_{b})^{2}$, 
$s_{1} = (p_{a} + q_{2})^{2} = (p_{1} + p_{34})^{2}$,
$s_{2} = (p_{b} + q_{1})^{2} = (p_{2} + p_{34})^{2}$,
$p_{34} = p_{3} + p_{4}$.
The amplitude for diffractive exclusive $\pi^+ \pi^-$ continuum production can be written as
the sum 
${\cal M}^{(\Pom \Pom \to \pi^{+}\pi^{-})} =
{\cal M}^{({\rm \hat{t}})}+{\cal M}^{({\rm \hat{u}})}$,
\begin{equation}
\begin{split}
{\cal M}^{({\rm \hat{t}})}_{\lambda_{a} \lambda_{b} \to \lambda_{1} \lambda_{2} \pi^{+}\pi^{-}} 
&= (-i)
\bar{u}(p_{1}, \lambda_{1}) 
i\Gamma^{(\Pom pp)}_{\mu_{1} \nu_{1}}(p_{1},p_{a}) 
u(p_{a}, \lambda_{a})\,
i\Delta^{(\Pom)\, \mu_{1} \nu_{1}, \alpha_{1} \beta_{1}}(s_{13},t_{1}) \,
i\Gamma^{(\Pom \pi\pi)}_{\alpha_{1} \beta_{1}}(\hat{p}_{t},-p_{3}) \,
i\Delta^{(\pi)}(\hat{p}_{t}) \\
& \quad \times  i\Gamma^{(\Pom \pi\pi)}_{\alpha_{2} \beta_{2}}(p_{4},\hat{p}_{t})\,
i\Delta^{(\Pom)\, \alpha_{2} \beta_{2}, \mu_{2} \nu_{2}}(s_{24},t_{2}) \,
\bar{u}(p_{2}, \lambda_{2}) 
i\Gamma^{(\Pom pp)}_{\mu_{2} \nu_{2}}(p_{2},p_{b}) 
u(p_{b}, \lambda_{b}) \,,
\end{split}
\label{amplitude_t}
\end{equation}
\begin{equation} 
\begin{split}
{\cal M}^{({\rm \hat{u}})}_{\lambda_{a} \lambda_{b} \to \lambda_{1} \lambda_{2} \pi^{+}\pi^{-}} 
&= (-i)\,
\bar{u}(p_{1}, \lambda_{1}) 
i\Gamma^{(\Pom pp)}_{\mu_{1} \nu_{1}}(p_{1},p_{a}) 
u(p_{a}, \lambda_{a}) \,
i\Delta^{(\Pom)\, \mu_{1} \nu_{1}, \alpha_{1} \beta_{1}}(s_{14},t_{1}) \,
i\Gamma^{(\Pom \pi\pi)}_{\alpha_{1} \beta_{1}}(p_{4},\hat{p}_{u}) 
\,i\Delta^{(\pi)}(\hat{p}_{u})  \\
& \quad \times  
i\Gamma^{(\Pom \pi\pi)}_{\alpha_{2} \beta_{2}}(\hat{p}_{u},-p_{3})\, 
i\Delta^{(\Pom)\, \alpha_{2} \beta_{2}, \mu_{2} \nu_{2}}(s_{23},t_{2}) \,
\bar{u}(p_{2}, \lambda_{2}) 
i\Gamma^{(\Pom pp)}_{\mu_{2} \nu_{2}}(p_{2},p_{b}) 
u(p_{b}, \lambda_{b}) \,,
\end{split}
\label{amplitude_u}
\end{equation}
where $\hat{p}_{t} = p_{a} - p_{1} - p_{3}$ and 
$\hat{p}_{u} = p_{4} - p_{a} + p_{1}$, $s_{ij} = (p_{i} + p_{j})^{2}$,
$\Delta^{(\pi)}(\hat{p}) = (\hat{p}^{2}-m_{\pi}^{2})^{-1}$.
For extensive discussions we refer to \cite{Ewerz:2013kda}.
The pomeron-proton vertex function, supplemented by a vertex form factor, 
taken here to be the Dirac electromagnetic form factor of the proton for simplicity,
has the form
%
\begin{eqnarray}
&&i\Gamma_{\mu \nu}^{(\Pom pp)}(p',p)= 
i\Gamma_{\mu \nu}^{(\Pom \bar{p} \bar{p})}(p',p)
\nonumber\\
&& \quad =-i 3 \beta_{\Pom NN} F_{1}\bigl((p'-p)^2\bigr)
\left\lbrace 
\frac{1}{2} 
\left[ \gamma_{\mu}(p'+p)_{\nu} 
     + \gamma_{\nu}(p'+p)_{\mu} \right]
- \frac{1}{4} g_{\mu \nu} ( p\!\!\!/' + p\!\!\!/ )
\right\rbrace\,,
\label{A4}
\end{eqnarray}
with 
$\beta_{\Pom NN} = 1.87$~GeV$^{-1}$.
For the $\Pom \pi \pi$ vertex we have
%
\begin{eqnarray}
i\Gamma_{\mu \nu}^{(\Pom \pi\pi)}(k',k)=
-i 2 \beta_{\Pom \pi\pi} 
\left[ (k'+k)_{\mu}(k'+k)_{\nu} - \frac{1}{4} g_{\mu \nu} (k' + k)^{2} \right] \, F_{M}((k'-k)^2)\,,
\label{vertex_pompipi}
\end{eqnarray}
with 
$\beta_{\Pom \pi\pi} = 1.76$~GeV$^{-1}$ and we use the pion electromagnetic form factor 
in a simple parametrization $F_{M}(\hat{p}^{2})= (1-\hat{p}^{2}/\Lambda_{0}^{2})^{-1}$, 
$\Lambda_{0}^{2} = 0.5$~GeV$^{2}$.
The off-shellness of the intermediate pions in (\ref{amplitude_t}) and (\ref{amplitude_u})
is taken into account by the inclusion of form factors. 
The form factors are parametrised in the monopole form $\hat{F}_{\pi}(\hat{p}^{2})=
(\Lambda^{2}_{off,M} - m_{\pi}^{2})/(\Lambda^{2}_{off,M} - \hat{p}^{2})$ or, alternatively, 
in the exponential form
$\hat{F}_{\pi}(\hat{p}^{2}) = \exp\left((\hat{p}^{2}-m_{\pi}^{2})/\Lambda^{2}_{off,E}\right)$,
where $\Lambda_{off,M}$ (or $\Lambda_{off,E}$) could be adjusted to experimental data.
Here the normalisation condition $\hat{F}_{\pi}(m_{\pi}^{2}) = 1$ is clearly satisfied.
%
%

Our effective pomeron propagator reads
\begin{eqnarray}
i \Delta^{(\Pom)}_{\mu \nu, \kappa \lambda}(s,t) = 
\frac{1}{4s} \left( g_{\mu \kappa} g_{\nu \lambda} 
                  + g_{\mu \lambda} g_{\nu \kappa}
                  - \frac{1}{2} g_{\mu \nu} g_{\kappa \lambda} \right)
(-i s \alpha'_{\Pom})^{\alpha_{\Pom}(t)-1}
\label{A1}
\end{eqnarray}
and fulfills the following relations:
\begin{eqnarray}
&&\Delta^{(\Pom)}_{\mu \nu, \kappa \lambda}(s,t) = 
\Delta^{(\Pom)}_{\nu \mu, \kappa \lambda}(s,t) =
\Delta^{(\Pom)}_{\mu \nu, \lambda \kappa}(s,t) =
\Delta^{(\Pom)}_{\kappa \lambda, \mu \nu}(s,t) \,, \nonumber \\ 
&&g^{\mu \nu} \Delta^{(\Pom)}_{\mu \nu, \kappa \lambda}(s,t) = 0, \quad 
g^{\kappa \lambda} \Delta^{(\Pom)}_{\mu \nu, \kappa \lambda}(s,t) = 0 \,.
\label{A2}
\end{eqnarray}
%
Here, the pomeron trajectory $\alpha_{\Pom}(t)$
is assumed to be of standard linear form, see e.g. \cite{Donnachie:2002en},
\begin{eqnarray}
\alpha_{\Pom}(t) = \alpha_{\Pom}(0)+\alpha'_{\Pom}\,t\,, \quad \alpha_{\Pom}(0) = 1.0808\,, \quad \alpha'_{\Pom} = 0.25 \; \mathrm{GeV}^{-2}\,.
\label{A3}
\end{eqnarray}

The $\Pom \Pom f_{2}$ vertex can be written as 
\begin{eqnarray}
i\Gamma_{\mu \nu,\kappa \lambda,\rho \sigma}^{(\Pom \Pom f_{2})} (q_{1},q_{2}) =
\left( i\Gamma_{\mu \nu,\kappa \lambda,\rho \sigma}^{(\Pom \Pom f_{2})(1)} \mid_{bare}
+ \sum_{j=2}^{7}i\Gamma_{\mu \nu,\kappa \lambda,\rho \sigma}^{(\Pom \Pom f_{2})(j)}(q_{1},q_{2}) \mid_{bare} 
\right)
\tilde{F}^{(\Pom \Pom f_{2})}(q_{1}^{2},q_{2}^{2},p_{34}^{2}) \,.
\label{vertex_pompomT}
\end{eqnarray}
A possible choice for the $\Pom \Pom f_{2}$ couplings, denoted by $j = 1, ..., 7$ terms, is given in Appendix~A of \cite{Lebiedowicz:2016ioh}.
Our attempts to determine the parameters of pomeron-pomeron-meson couplings
as far as possible from experimental data have been presented in 
\cite{Lebiedowicz:2013ika,Lebiedowicz:2016ioh,Lebiedowicz:2018eui}.
Other details as form of form factors, the tensor-meson propagator $\Delta^{(f_{2})}$ and
the $f_{2} \pi \pi$ vertex are given in \cite{Ewerz:2013kda,Lebiedowicz:2016ioh,Lebiedowicz:2018eui}.

The ansatz for the $C=+1$ reggeons $\Reg_{+} = f_{2 \Reg}, a_{2 \Reg}$
is similar to (\ref{A4}) - (\ref{A3}).
The $f_{2 \Reg}$- and $a_{2 \Reg}$-proton vertex functions are obtained from (\ref{A4})
with the replacements ($M_{0} = 1$~GeV)
$3 \beta_{\Pom NN} \to \frac{g_{f_{2 \Reg} pp}}{M_{0}}$,
$g_{f_{2 \Reg} pp} = 11.04$,
and
$3 \beta_{\Pom NN} \to \frac{g_{a_{2 \Reg} pp}}{M_{0}}$,
$g_{a_{2 \Reg} pp} = 1.68$, respectively.
The $f_{2 \Reg}$-pion vertex function is obtained from (\ref{vertex_pompipi})
with the replacement $2 \beta_{\Pom \pi\pi} \to \frac{g_{f_{2 \Reg} \pi\pi}}{2 M_{0}}$, 
$g_{f_{2 \Reg} \pi\pi} = 9.30$.
The $\Reg_{+}$ propagator is obtained from (\ref{A1}) with the replacements
$\alpha_{\Pom}(t) \to \alpha_{\Reg_{+}}(t) = \alpha_{\Reg_{+}}(0)+\alpha'_{\Reg_{+}}t$,
$\alpha_{\Reg_{+}}(0) = 0.5475$,
$\alpha'_{\Reg_{+}} = 0.9 \; \mathrm{GeV}^{-2}$.
The $\Reg_{-}$-proton vertex (for the $C=-1$ reggeons $\Reg_{-} = \omega_{\Reg}, \rho_{\Reg}$) reads 
\begin{eqnarray}
i\Gamma_{\mu}^{(\Reg_{-} pp)}(p',p)= -i\Gamma_{\mu}^{(\Reg_{-} \bar{p} \bar{p})}(p',p)
                                   = -i g_{\Reg_{-} pp} F_{1}\bigl((p'-p)^2\bigr) \gamma_{\mu}\,,
\label{A10}
\end{eqnarray}
with $g_{\omega_{\Reg} pp} = 8.65$ and $g_{\rho_{\Reg} pp} = 2.02$.
For the $\rho_{\Reg}$-pion vertex we write
\begin{eqnarray}
i\Gamma_{\mu}^{(\rho_{\Reg} \pi^{+}\pi^{+})}(k',k)
= -i\Gamma_{\mu}^{(\rho_{\Reg} \pi^{-}\pi^{-})}(k',k)
= -\frac{i}{2} g_{\rho_{\Reg} \pi\pi} F_{M}\bigl((k'-k)^2\bigr) (k'+k)_{\mu}\,,
\label{A12}
\end{eqnarray}
with $\rho_{\Reg} \pi \pi = 15.63$.
We assume an effective vector propagator
\begin{eqnarray}
i \Delta^{(\Reg_{-})}_{\mu \nu}(s,t) = 
i g_{\mu \nu} \frac{1}{M_{-}^{2}} (-i s \alpha'_{\Reg_{-}})^{\alpha_{\Reg_{-}}(t)-1}\,,
\label{A8}
\end{eqnarray}
with
$\alpha_{\Reg_{-}}(t) = \alpha_{\Reg_{-}}(0)+\alpha'_{\Reg_{-}}t$,
$\alpha_{\Reg_{-}}(0) = 0.5475$,
$\alpha'_{\Reg_{-}} = 0.9 \; \mathrm{GeV}^{-2}$,
and $M_{-} = 1.41 \; \mathrm{GeV}$.

Our ansatz for the odderon follows (3.16), (3.17) and (3.68), (3.69) of \cite{Ewerz:2013kda}:
\begin{eqnarray}
&&i \Delta^{(\Ode)}_{\mu \nu}(s,t) = 
-i g_{\mu \nu} \frac{\eta_{\Ode}}{M_{0}^{2}} (-i s \alpha'_{\Ode})^{\alpha_{\Ode}(t)-1}\,,
\label{A13} \\
&&i\Gamma_{\mu}^{(\Ode pp)}(p',p) = -i\Gamma_{\mu}^{(\Ode \bar{p} \bar{p})}(p',p)
                                  = -i 3\beta_{\Ode pp} M_{0}\,F_{1}\bigl((p'-p)^2\bigr) \gamma_{\mu}\,.
\label{A14}
\end{eqnarray}
We take here what we think are representative values
for the odderon parameters in light of the recent TOTEM results \cite{Antchev:2017yns},
$\eta_{\Ode} = -1$,
$\alpha_{\Ode}(t) = \alpha_{\Ode}(0)+\alpha'_{\Ode} t$,
$\alpha_{\Ode}(0) = 1.05,\, \alpha'_{\Ode} = 0.25 \mathrm{GeV}^{-2}$,
$\beta_{\Ode NN} = 0.2 \; \mathrm{GeV}^{-1}$.
All numbers for the parameters listed above should be considered as default values to be checked
and -- if necessary -- adjusted using relevant experimental data.
Some estimates of the present uncertainties of the parameters
are discussed in Sec.~3 of Ref.~\cite{Ewerz:2013kda}.

In reality the Born approximation, e.g. for the amplitude (\ref{amplitude_f2_pomTpomT}), is not sufficient and 
absorption corrections (rescattering effects) must be taken into account,
see \cite{Harland-Lang:2013dia,Lebiedowicz:2015eka}.
A Monte Carlo generator containing a various processes 
and including detector effects (acceptance, efficiency)
would be useful in theory-data comparison.
The $\mathtt{GenEx}$ Monte Carlo generator \cite{Kycia:2014hea} 
could be used in this context.

\section{Selected results}
\label{results}

\subsection{$pp$ and $\bar{p}p$ elastic scattering}
\label{sec-1}
\begin{figure}[h]
\centering
\includegraphics[width=5.5cm,clip]{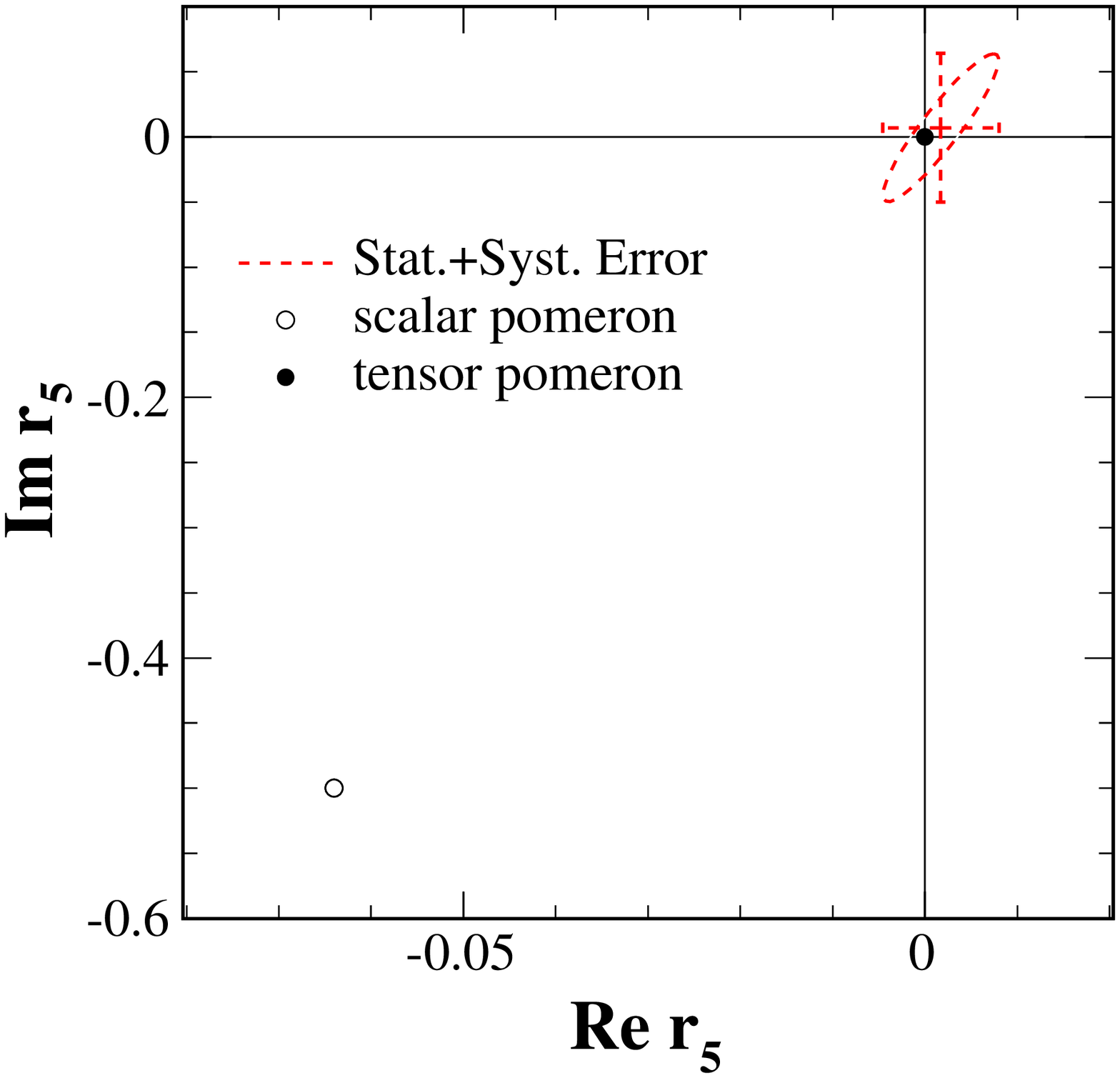}
\includegraphics[width=5.5cm,clip]{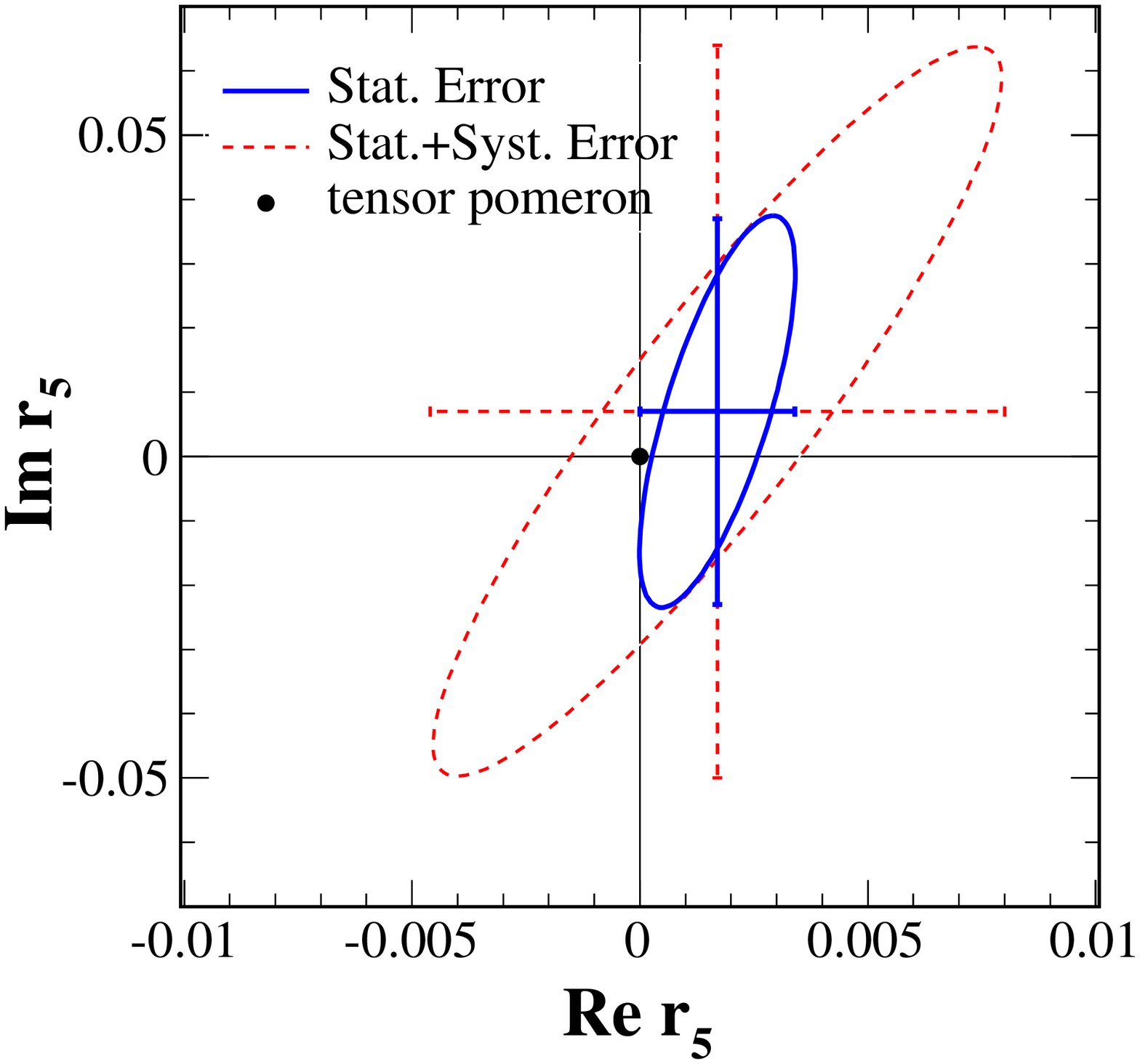}
\caption{The experimental results for $r_{5}$ at $\sqrt{s} = 200$~GeV
from Fig.~5 of \cite{Adamczyk:2012kn} together with the results for
the tensor and the scalar pomeron from \cite{Ewerz:2016onn}.}
\label{fig:pp_pp}
\end{figure}

In \cite{Ewerz:2016onn} we confronted three hypotheses for the soft pomeron,
tensor, vector, and scalar, with current experimental data on polarised $pp$ elastic scattering \cite{Adamczyk:2012kn}.
For the vector-pomeron case a big problem arise if we consider $pp$ and $\bar{p}p$ scattering.
Taken literally it gives opposite signs for $pp$ and $\bar{p}p$ total cross sections.
Thus, we shall not consider a vector pomeron further.
In order to discriminate between the tensor and scalar cases
we turn to the experiment \cite{Adamczyk:2012kn}.
There a good measurement of the ratio
of single-flip to non-flip amplitudes at $\sqrt{s} = 200$~GeV and 
for $0.003 \leqslant |t| \leqslant 0.035$~GeV$^{2}$ was performed.
The relevant quantity is
%
$r_{5}(s,t) = 
\frac{2m_{p} \,\phi_{5}(s,t)}{\sqrt{-t} \,
{\rm Im}\left[ \phi_{1}(s,t) + \phi_{3}(s,t)\right]}$.
%
The complex $r_{5}$ parameter is only weakly dependent on $t$.
Therefore, we can approximately set $t=0$
and get
$r_{5}^{\Pom_{T}}(s,0) = \left( -0.28 - i 2.20 \right) \times 10^{-5}$ 
and $r_{5}^{\Pom_{S}}(s,0) = -0.064 - i 0.500$
for the scalar- and tensor-pomeron model, respectively.
In Fig.~\ref{fig:pp_pp} we show the STAR experimental result \cite{Adamczyk:2012kn} 
together with the scalar-pomeron and tensor-pomeron results.
Clearly, the tensor-pomeron result is perfectly
compatible with the experiment whereas the scalar-pomeron result
is far outside the experimental error ellipse.

\subsection{$pp \to pp \pi^{+} \pi^{-}$ and $pp \to pp K^{+} K^{-}$ reactions}
\label{sec-2}

\begin{figure}[h]
\centering
\includegraphics[width=5.8cm,clip]{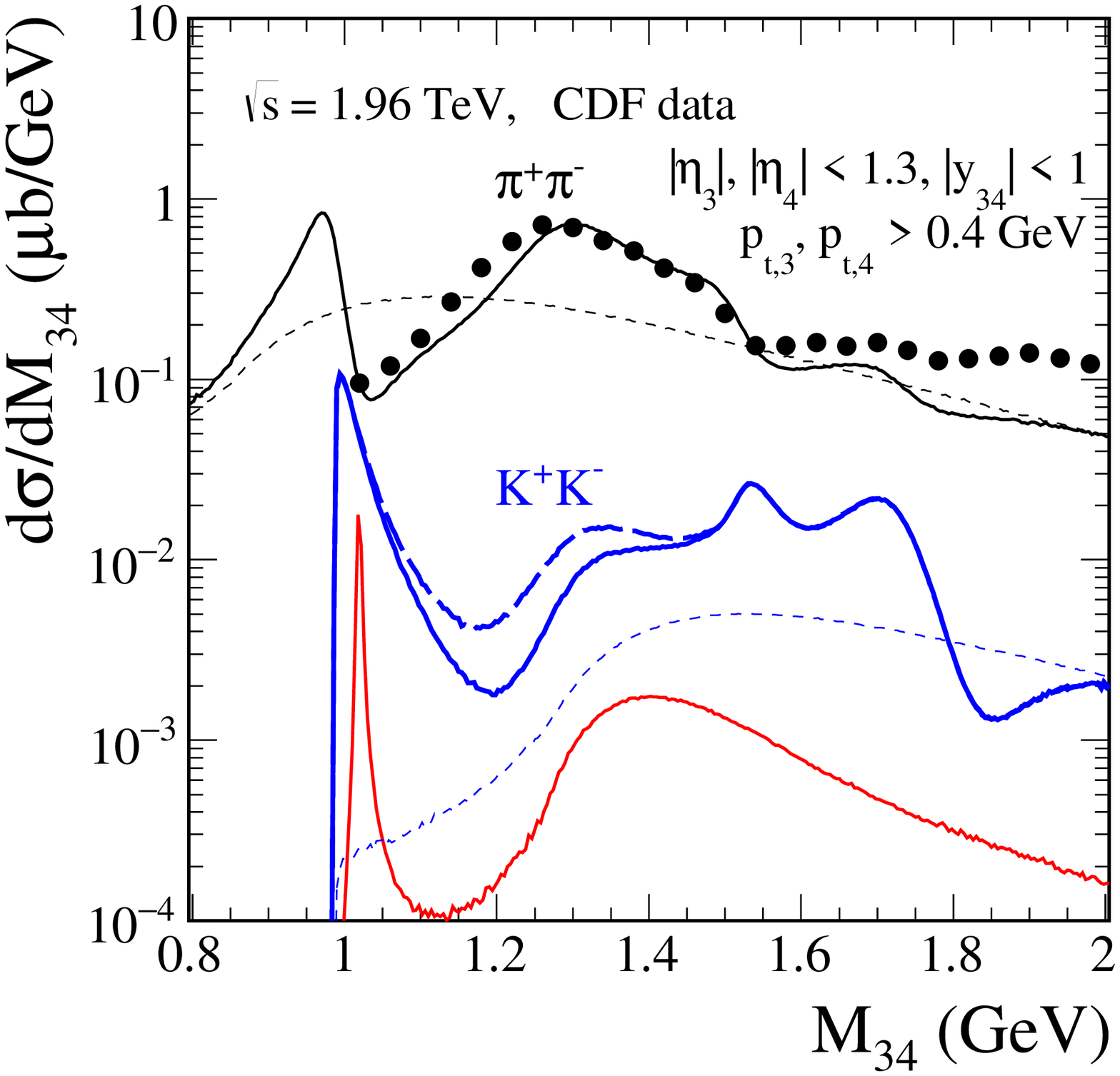}
\includegraphics[width=5.8cm,clip]{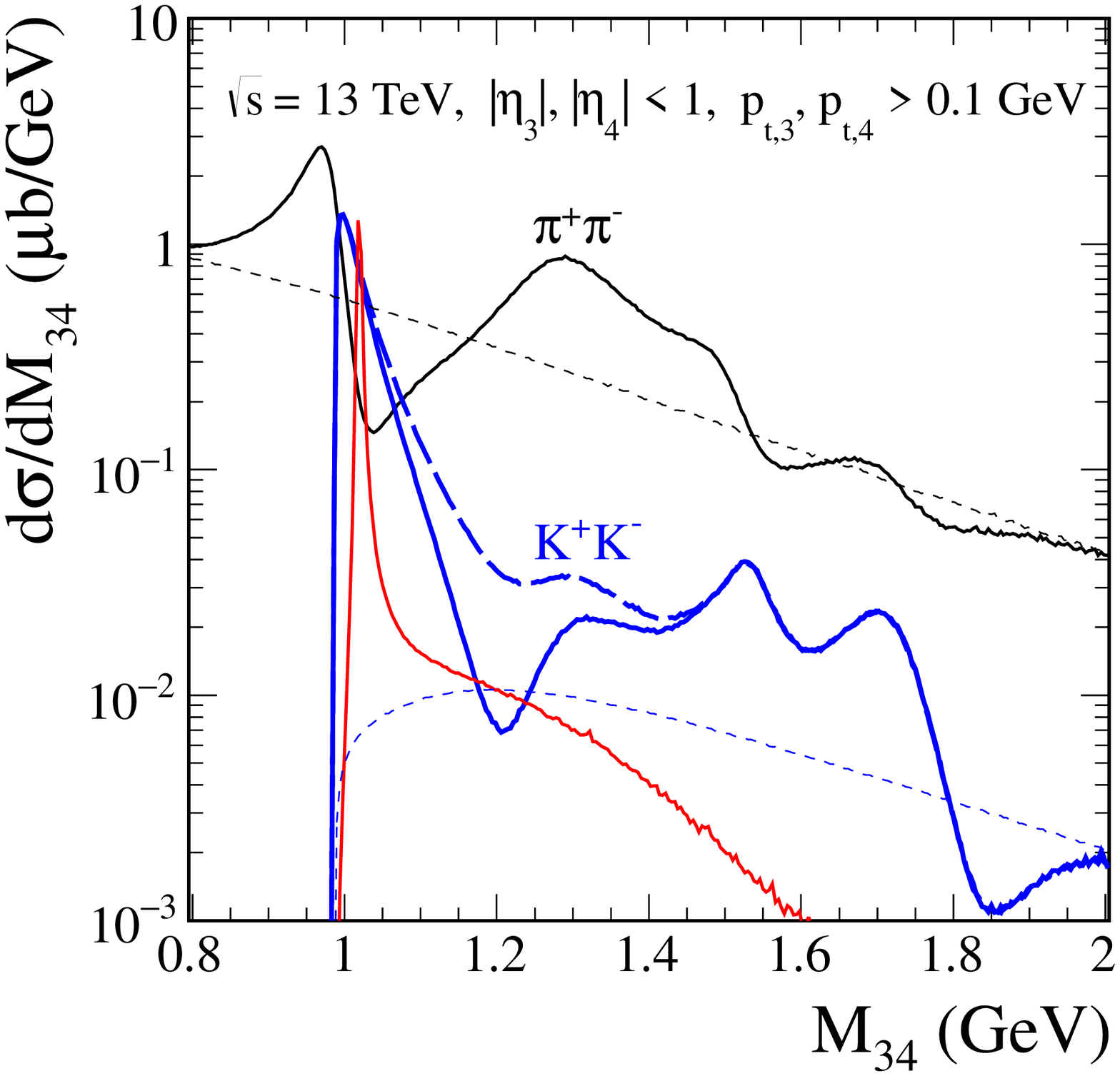}
\caption{The invariant mass distributions for
centrally produced $\pi^{+} \pi^{-}$ (the black top lines) 
and $K^{+} K^{-}$ (the blue bottom lines) pairs
with the relevant experimental kinematical cuts specified in the legend.
Theoretical results including both the non-resonant continuum and resonances are represented
by the solid lines, respectively.
The short-dashed lines represent the purely diffractive continuum term alone.
The CDF experimental data from \cite{Aaltonen:2015uva} in the left panel
for the $p\bar{p} \to p\bar{p} \pi^{+} \pi^{-}$ reaction are presented.
For the $pp \to ppK^{+}K^{-}$ reaction the solid and long-dashed blue lines correspond to 
the results for $\phi_{f_{0}(980)} = 0$ and $\pi/2$ in the coupling 
constant $g_{f_{0}(980) K^{+} K^{-}} \,e^{i \phi_{f_{0}(980)}}$, respectively.
The lower red line represents the $\phi(1020)$ meson 
plus continuum photoproduction contribution.
Absorption effects were taken into account effectively by the gap survival factors,
$\langle S^{2} \rangle = 0.1$ for the purely diffractive contributions
and $\langle S^{2} \rangle = 0.9$ for the photoproduction contributions.}
\label{fig:dsig_dM34}
\end{figure}

In Fig.~\ref{fig:dsig_dM34} we present the dipion/dikaon invariant mass distributions 
imposing experimental cuts.
The short-dashed lines represent the purely diffractive continuum term 
including both pomeron and reggeon exchanges,
discussed in \cite{Lebiedowicz:2016ioh,Lebiedowicz:2018eui}.
Exclusive production of light mesons 
both in the $pp \to pp\pi^{+}\pi^{-}$ and $pp \to ppK^{+}K^{-}$ reactions
are measurable at RHIC and LHC.
The pattern of visible structures in the invariant mass distributions
is related to the scalar and tensor isoscalar mesons
and it depends on experimental kinematics.
For example, for the $pp \to ppK^{+}K^{-}$ reaction presented in Fig.~\ref{fig:dsig_dM34},
the solid lines represent the coherent sum of the diffractive
continuum, and the scalar $f_{0}(980)$, $f_{0}(1500)$, $f_{0}(1710)$,
and tensor $f_{2}(1270)$, $f'_{2}(1525)$ resonances.
The lower red lines show the photoproduction term including
the dominant $\phi(1020) \to K^{+}K^{-}$ and the continuum (Drell-S\"oding) contributions.
The narrow $\phi$ resonance is visible above the continuum term.
It may, in principle, be visible on top of the broader $f_{0}(980)$ resonance.
The coupling parameters of the tensor pomeron to the $\phi$ meson was
fixed based on the HERA experimental data for the $\gamma p \to \phi p$ reaction.

One can expect, with our default choice of parameters,
that the scalar $f_{0}(980)$, $f_{0}(1500)$, $f_{0}(1710)$
and the tensor $f_{2}(1270)$, $f'_{2}(1525)$ mesons
will be easily identified experimentally in CEP processes.
The absorption effects lead to a huge damping of the cross section for the purely
diffractive contribution and a relatively small reduction
of the cross section for the $\phi(1020)$ photoproduction contribution.
Therefore we expect that one could observe the $\phi$ resonance term,
especially when no restrictions on the leading protons are included.
We note that central exclusive production of $\phi$ 
offers also the possibility to search for
effects of the elusive odderon, as was pointed out in \cite{Schafer:1991na}.

\begin{figure}[h]
\centering
(a)\includegraphics[width=5.8cm,clip]{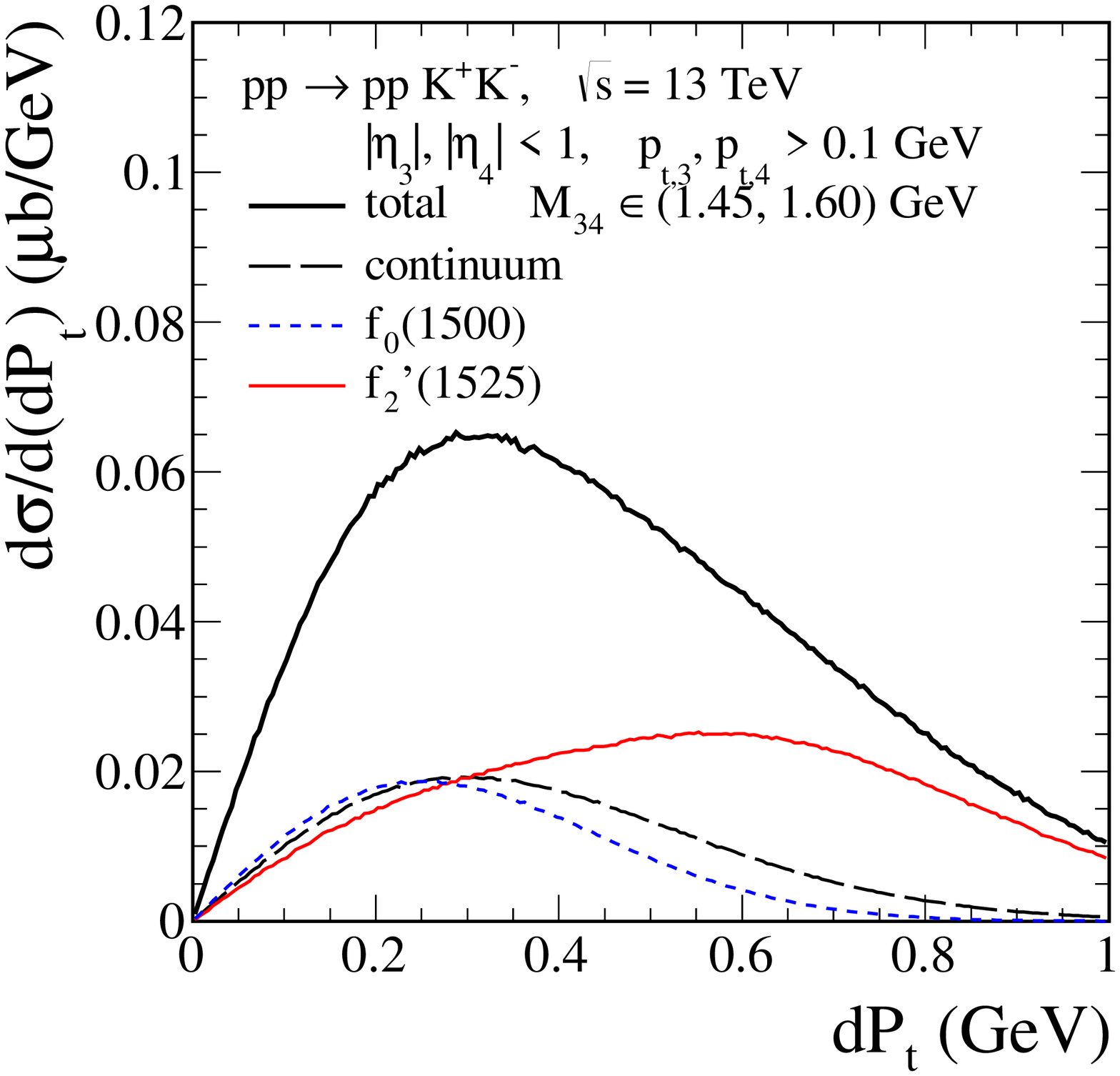}
(b)\includegraphics[width=5.8cm,clip]{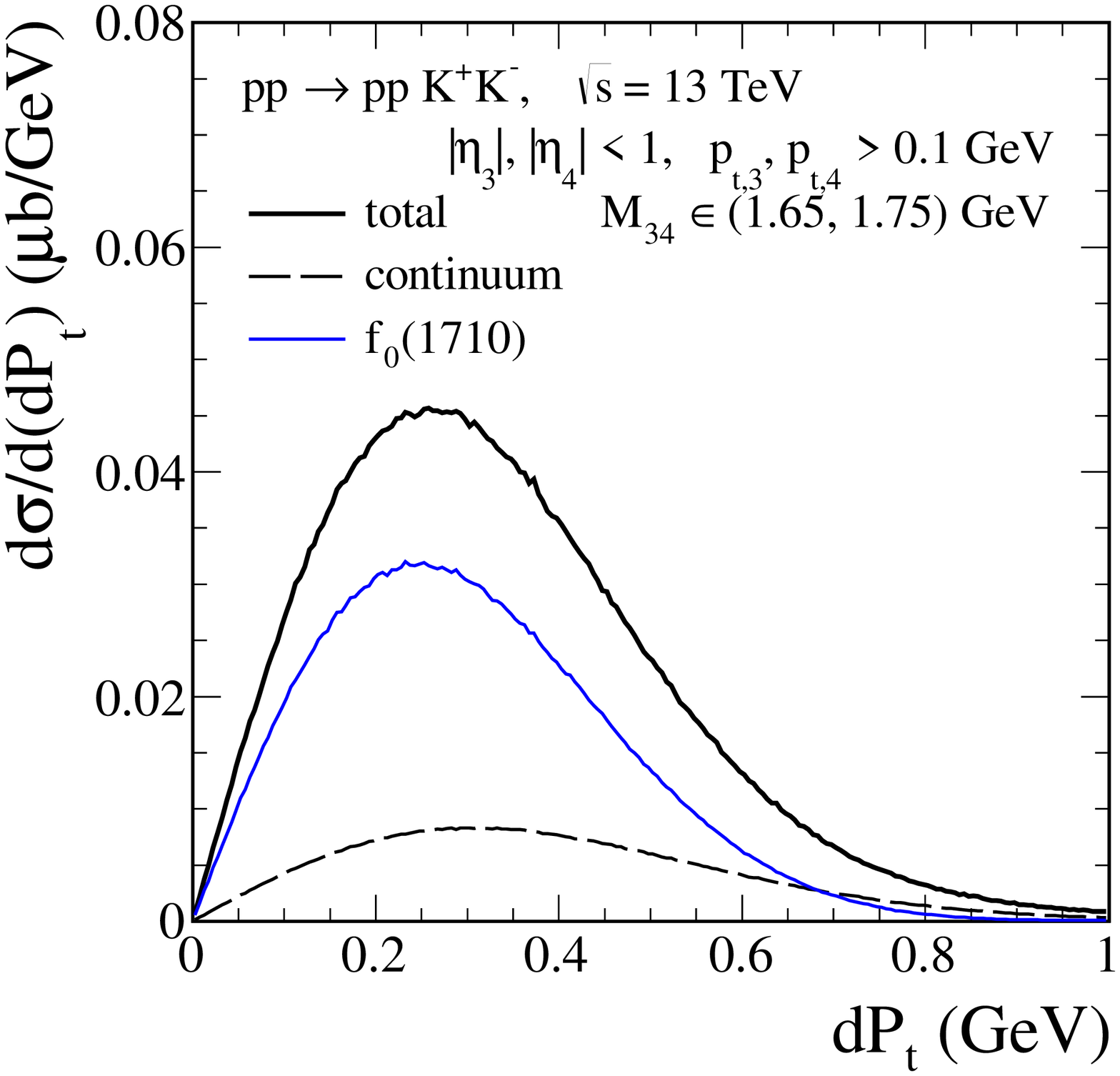}
\caption{
The differential cross sections $d\sigma/d(dP_{t})$
as a function of the $dP_{t}$ ``glueball filter'' variable
for the $pp \to pp K^{+} K^{-}$ reaction.
Calculations were done for $\sqrt{s} = 13$~TeV, $|\eta_{K}| < 1$,
$p_{t,K} > 0.1$~GeV, and in two dikaon invariant mass regions:
(a) $M_{34} \in (1.45, 1.60)$~GeV and (b) $M_{34} \in (1.65, 1.75)$~GeV.}
\label{fig:dsig_dptperp}
\end{figure}

In Fig.~\ref{fig:dsig_dptperp} we present distributions 
in a special "glueball filter variable" $dP_{t}$ \cite{Close:1997pj}
defined by the difference of the transverse momentum vectors $dP_{t} = |\bdPt|$,
$\bdPt = \bqta - \bqtb = \bptb - \bpta$.
Results for the ALICE kinematics and for two $M_{K^{+}K^{-}}$ regions are shown.
No absorption effects were taken into account here.
We see that the maximum for the $q \bar{q}$ state $f'_{2}(1525)$
is around $dP_{t} = 0.6$~GeV. On the other hand,
for the scalar glueball candidates $f_{0}(1500)$ and $f_{0}(1710)$
the maximum is around $dP_{t} = 0.25$~GeV, that is,
at a lower value than for the $f'_{2}(1525)$ resonance.
This is also in accord with the discussion in Refs.~\cite{Barberis:1996iq,Barberis:1999cq}.

\subsection{$pp \to pp p \bar{p}$ reaction}
\label{sec-3}
\begin{figure}[h]
\centering
\includegraphics[width=5.8cm,clip]{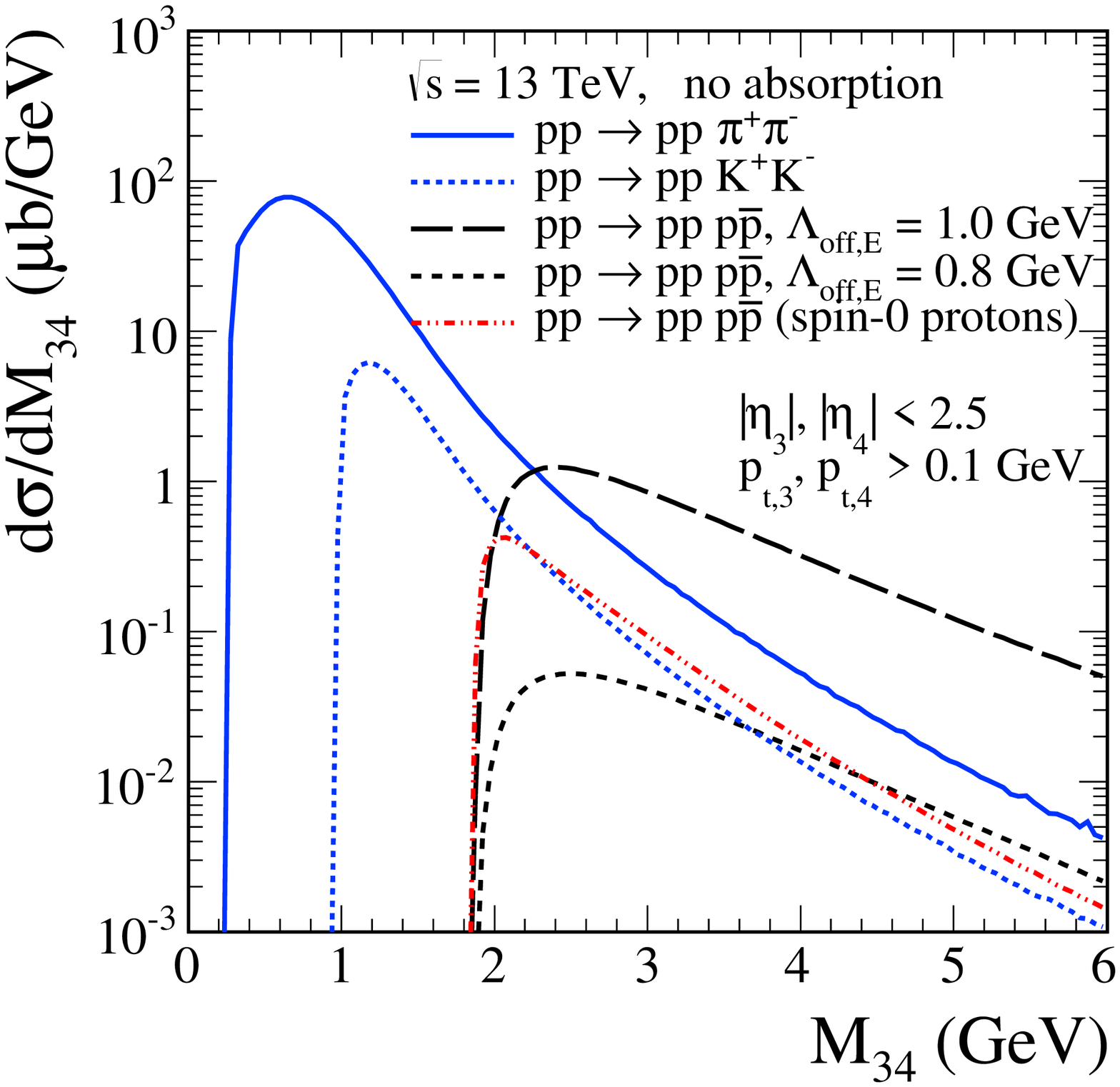}
\includegraphics[width=5.8cm,clip]{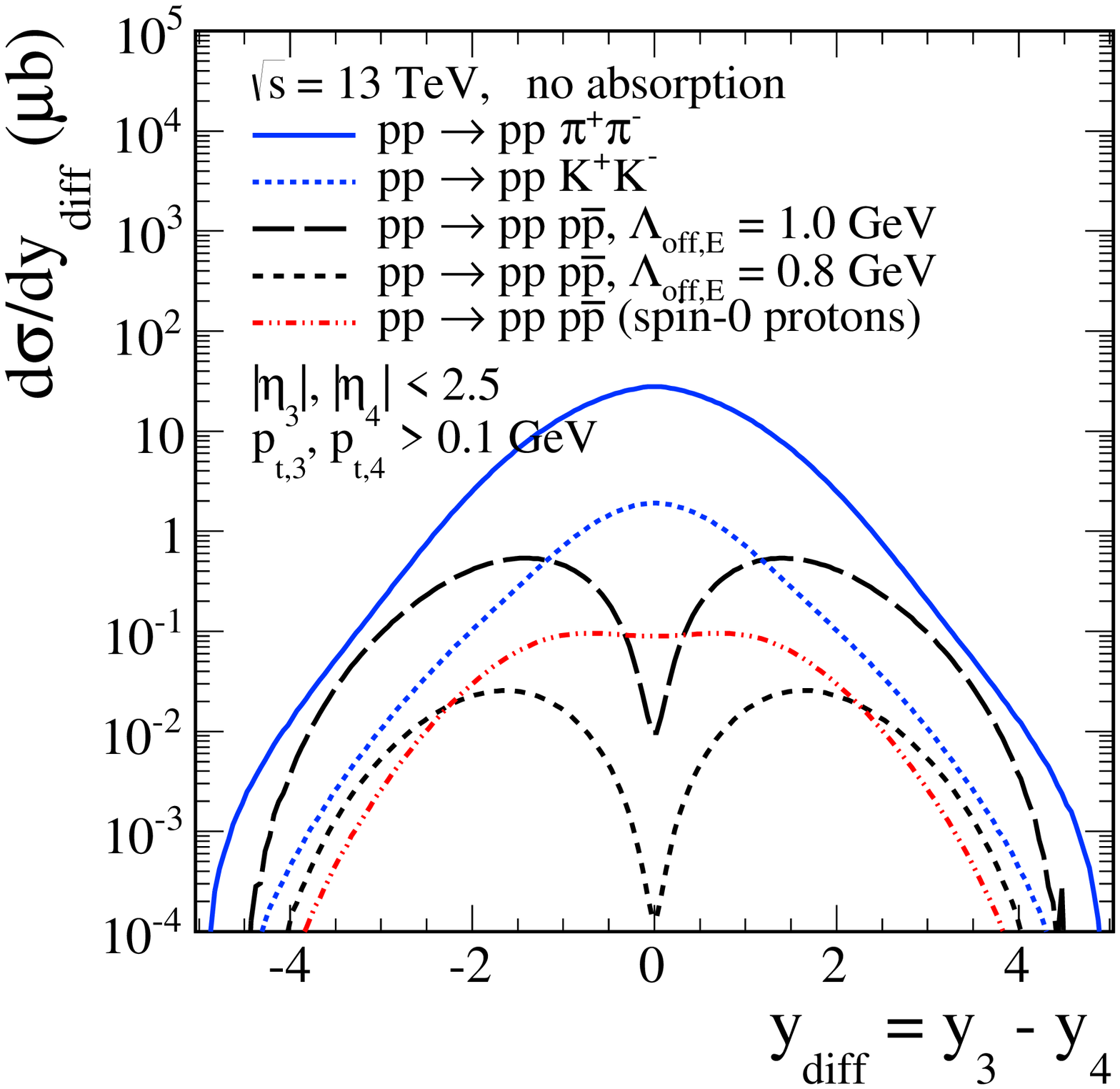}
\caption{
In the left panel we show the invariant mass distributions for 
centrally produced $\pi^+ \pi^-$, $K^+ K^-$ and $p \bar{p}$ systems
for a typical LHC experimental conditions at $\sqrt{s} = 13$~TeV.
Results for the combined tensor-pomeron and reggeon exchanges
and $\Lambda_{off,E} = 1$~GeV are presented.
For the $p \bar{p}$ production we show results 
also for $\Lambda_{off,E} = 0.8$~GeV.
In the right panel we show the distributions in the rapidity difference 
between the centrally produced hadrons.
No absorption effects were included here.}
\label{fig:ppbar}
\end{figure}

In Fig.~\ref{fig:ppbar}, we show the invariant mass distributions for 
centrally produced $\pi^+ \pi^-$, $K^+ K^-$ and $p \bar{p}$ systems (the left panel)
and the distributions in rapidity difference 
${\rm y}_{diff} = {\rm y}_{3} - {\rm y}_{4}$ (the right panel)
at $\sqrt{s} = 13$~TeV.
In our calculations we include both the tensor-pomeron and the reggeon $\Reg_{+}$ and $\Reg_{-}$ exchanges. 
We predict a dip in the rapidity difference between 
the antiproton and proton for ${\rm y}_{diff} = 0$.
This novel effect is inherently related 
to the spin 1/2 of the produced hadrons.
We have checked that for the $p \bar{p}$ production
the ${\rm \hat{t}}$- and ${\rm \hat{u}}$-channel
diagrams interfere destructively for
$(C_{1},C_{2}) = (1,1)$ and $(-1,-1)$ exchanges
and constructively for $(1,-1)$ and $(-1,1)$ exchanges.
For the $\pi^{+} \pi^{-}$ production,
we get the opposite interference effects between
the ${\rm \hat{t}}$- and ${\rm \hat{u}}$-channel diagrams.

\begin{figure}[h]
\centering
\includegraphics[width=5.8cm,clip]{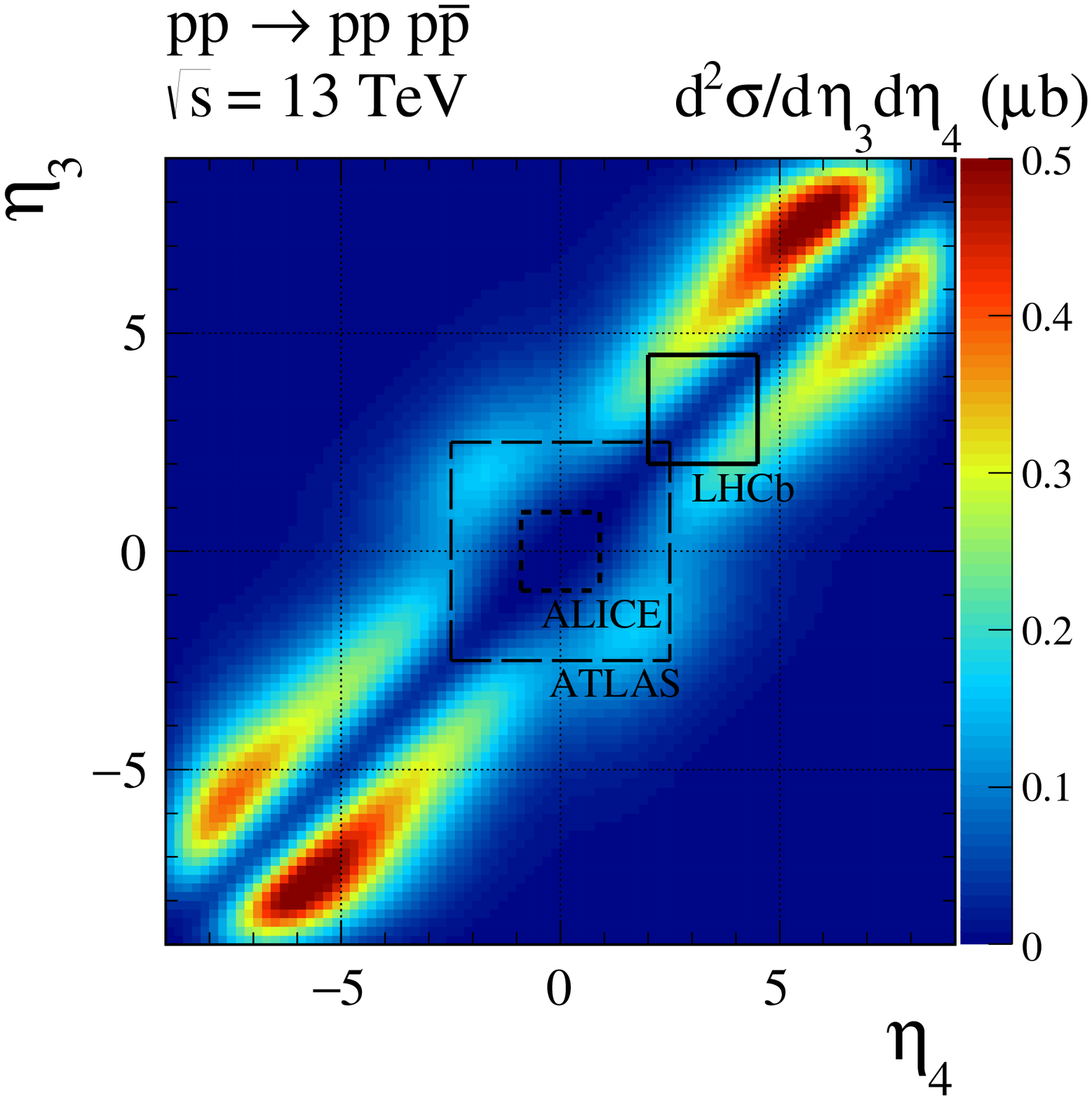}
\includegraphics[width=5.8cm,clip]{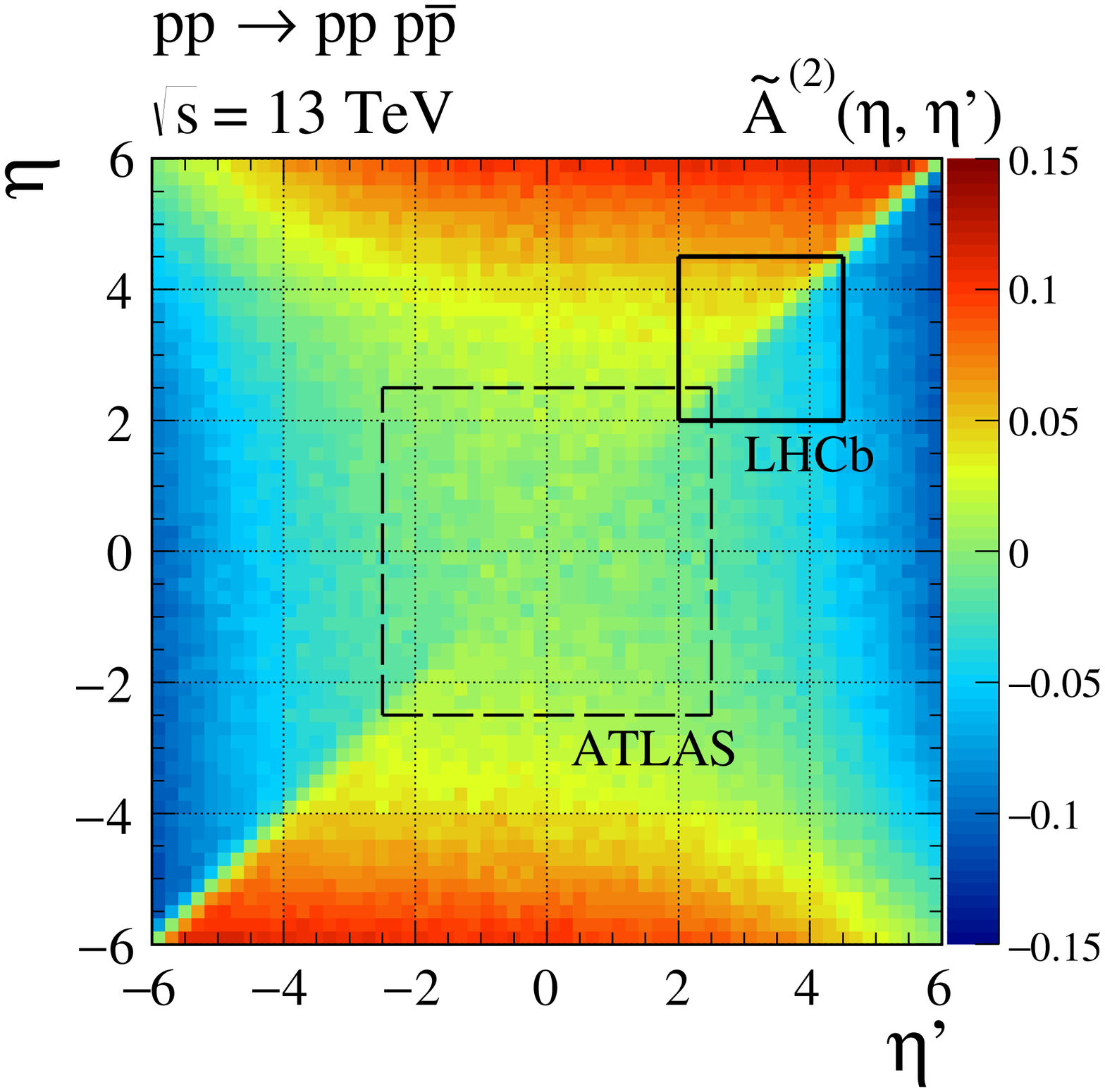}
\caption{
The two-dimensional distribution in ($\eta_{3}, \eta_{4}$) 
for the diffractive continuum $p\bar{p}$ production for the full phase space
and the asymmetry $\widetilde{A}^{(2)}(\eta,\eta')$ [see Eq.~(\ref{asymmetry_3})]
at $\sqrt{s} = 13$~TeV.
In addition, regions of the coverage for the ALICE, ATLAS and LHCb
experiments are shown.
}
\label{fig:5b}
\end{figure}

In our calculations we have included both pomeron and reggeon exchanges. 
The reggeon exchange contributions lead to an enhancement of the cross section
mostly at large rapidities of the centrally produced hadrons.
For the production of the dipion continuum, the cross section 
is concentrated along the diagonal $\eta_{3} = \eta_{4}$.
For the production of $p \bar{p}$ pairs, the dip extends over the whole
diagonal in ($\eta_{3},\eta_{4}$) space, see the left panel in Fig.~\ref{fig:5b}.
In the right panel of Fig.~\ref{fig:5b} we show the asymmetry defined
for two pseudorapidities $\eta$ and $\eta'$
\begin{equation}\label{asymmetry_3}
\begin{split}
\widetilde{A}^{(2)}(\eta,\eta')=\frac{\frac{d^{2}\sigma}{d\eta_{3} d\eta_{4}}(\eta,\eta')
                                 -\frac{d^{2}\sigma}{d\eta_{3} d\eta_{4}}(\eta',\eta)}
                                 {\frac{d^{2}\sigma}{d\eta_{3} d\eta_{4}}(\eta,\eta')
                                 +\frac{d^{2}\sigma}{d\eta_{3} d\eta_{4}}(\eta',\eta)}\,.
\end{split}
\end{equation}
For the investigated pseudorapidity range the asymmetries 
due to pomeron plus reggeon exchanges show a positive sign for $|\eta| > |\eta'|$ 
and negative sign for $|\eta| < |\eta'|$.
That is, antiprotons are predicted to come out typically
with a higher absolute value of the (pseudo)rapidity than protons.
The asymmetry is caused by interference effects 
of the dominant $(\Pom, \Pom)$ with the subdominant 
$(\Reg_{-}, \Pom + \Reg_{+})$ and $(\Pom + \Reg_{+}, \Reg_{-})$ exchanges.
We have checked that in the limited range of pseudorapidities corresponding 
to the ATLAS and LHCb experiments
the effects of the secondary reggeons are predicted to be 
in the ranges of 2 - 11~\% and 5 - 26~\%, respectively.
The addition of an odderon with the parameters of (\ref{A13}) \textit{et seq.}
has only an effect of less than 0.5~\%.

In \cite{Lebiedowicz:2018sdt} we have discussed a first qualitative attempt to ``reproduce''
the experimentally observed behavior of the $p \bar{p}$ invariant mass ($M_{34}$) spectra
observed in \cite{Breakstone:1989ty,Barberis:1998sr}.
Our calculation shows that the diffractive production of $p \bar{p}$ 
through the $s$-channel $f_{0}(2100)$ resonance
leads to an enhancement at low $M_{34}$ and
that the resonance contribution is concentrated at $|{\rm y}_{diff}| < 1$.
In general, more resonances can contribute, 
e.g., $f_{0}(2020)$, $f_{0}(2200)$, $f_{0}(2300)$, $f_{2}(1950)$.
Also, the subthreshold $m_R < 2 m_p$ resonances that 
would effectively generate a continuum $p \bar{p}$ contribution
should be taken into account; see \cite{Klusek-Gawenda:2017lgt}.
Interference effects between the continuum and resonant mechanisms 
certainly will occur. We leave this interesting issue for future studies.

\section{Conclusions}
\label{conclusions}

When considering $pp$ and $\bar{p}p$ elastic scattering and the ratio
of helicity-flip to non-flip amplitudes 
we found that only the tensor pomeron, introduced in \cite{Ewerz:2013kda},
is a viable option for the soft pomeron.
In terms of elementary exchanges this should be viewed
as a coherent sum of exchanges of spin $2+4+6+...$.
This is the structure obtained in \cite{Nachtmann:1991ua} within
nonperturbative QCD investigations.
Investigations of the pomeron using the models of AdS/QCD
also prefer a tensor nature for pomeron exchange \cite{Iatrakis:2016rvj}.

We have given a consistent treatment of continuum and resonance production for different processes 
in central exclusive $pp$ and $p\bar{p}$ collisions
in an effective field-theoretic approach.
We have analysed the central exclusive production of $\pi^{+}\pi^{-}$ and $K^{+}K^{-}$ pairs at high energies. 
We have taken into account purely diffractive and diffractive photoproduction mechanisms.
For the purely diffractive mechanism we have included
the continuum and the dominant scalar $f_{0}(980)$, $f_{0}(1500)$, $f_{0}(1710)$
and tensor $f_{2}(1270)$, $f'_{2}(1525)$ resonances.
The amplitudes have been calculated using Feynman rules within
the tensor-pomeron model \cite{Ewerz:2013kda}.
The effective Lagrangians and the vertices for $\Pom \Pom$ fusion 
into the scalar and tensor mesons were discussed in
\cite{Lebiedowicz:2013ika,Lebiedowicz:2016ioh,Lebiedowicz:2018eui}.
Some model parameters ($\Pom \Pom M$ couplings, 
the off-shell dependence of form factors)
have been roughly adjusted to CDF data \cite{Aaltonen:2015uva}
and then used for predictions for the STAR, ALICE, ATLAS, CMS and LHCb experiments.
The distributions, in the so-called glueball filter variable $dP_{t}$, 
show different behavior in the $K^{+}K^{-}$ invariant mass windows 
around glueball candidates with masses $\sim 1.5$~GeV and $\sim 1.7$~GeV
than in other regions. The $dP_{t}$ distribution may help to interpret the relative rates
between the $f_{0}$ and $f_{2}$ resonances.

The photoproduction and purely diffractive 
contributions have different dependences on the proton transverse momenta.
Furthermore, the absorptive corrections for the photoproduction processes lead
to a much smaller reduction of the cross section than for the diffractive ones.
It can therefore be expected that the $\rho$- and $\phi$-photoproduction 
will be seen in experiments requiring only a very small deflection angle
for at least one of the outgoing protons.
However, we must keep in mind that other processes can contribute in experimental studies 
of exclusive photoproduction where only large rapidity gaps 
around the centrally produced vector meson are checked 
and the forward and backward going protons are not detected.
Experimental results for this kind of processes were published 
by the CDF \cite{Aaltonen:2015uva} and CMS \cite{Khachatryan:2017xsi} collaborations. 
We refer the reader to Ref.~\cite{Lebiedowicz:2016ryp} in which
$\rho^{0}$ production in $pp$ collisions was studied with one proton
undergoing diffractive excitation to a $\pi N$ system. 

Recently, in \cite{Lebiedowicz:2018sdt} we discussed
exclusive production of $p \bar{p}$ and $\Lambda \overline{\Lambda}$
pairs in proton-proton collisions. 
At the present stage, we have taken into account 
mainly the diffractive production of the $p \bar{p}$ continuum.
For our predictions for the LHC we have used 
the off-shell proton form factor parameter in the range
0.8~GeV $<\Lambda_{off,E}<$~1 GeV.
The invariant mass distribution for $p \bar{p}$ pairs is predicted to
extend to larger dihadron invariant masses
than for the production of $\pi^+ \pi^-$ or $K^+ K^-$.
Especially interesting is the distribution in the rapidity difference
between centrally produced antiproton and proton.
For continuum $p \bar{p}$ production, we predict a dip at ${\rm y}_{diff} = 0$, 
in contrast to $\pi^+ \pi^-$ and $K^+ K^-$ production 
in which a maximum of the cross section occurs at ${\rm y}_{diff} = 0$.
The dip is caused by a good separation of ${\rm \hat{t}}$ and ${\rm \hat{u}}$ contributions 
in ($\eta_{3}, \eta_{4}$) space and destructive interference of them
along the diagonal $\eta_{3} = \eta_{4}$ 
characteristic for our Feynman diagrammatic 
calculation with correct treatment of spins.
Any experimentally observed distortions from our continuum-$p \bar{p}$ predictions 
may therefore signal the presence of resonances. 
This could give new interesting information for meson spectroscopy.

To describe the relatively low-energy 
ISR and WA102 data \cite{Breakstone:1989ty,Barberis:1998sr} for the $pp \to pp p\bar{p}$ process
we find that we must include also subleading reggeon exchanges 
in addition to the two-pomeron exchange.
Then we made predictions for the LHC energy.
The reggeon exchange contributions 
lead to enhancements at large absolute values of the $p$ and $\bar{p}$ (pseudo)rapidities, 
see Fig.~\ref{fig:5b}.
A similar effect was predicted 
for the $pp \to pp \pi^{+} \pi^{-}$ reaction in \cite{Lebiedowicz:2010yb}.
We have predicted asymmetries in the (pseudo)rapidity distributions of
the centrally produced antiproton and proton.
The asymmetry should be much more visible for the LHCb experiment
which covers a region of larger pseudorapidities 
where the reggeon exchanges become more relevant.
Also the odderon will contribute to such asymmetries.
However, we find for typical odderon parameters 
allowed by recent $pp$ elastic data \cite{Antchev:2017yns}
only very small effects,
roughly a factor 10 smaller than the effects due to secondary reggeons.

\begin{acknowledgement}
This work was supported by Polish Ministry of Science and Higher Education 
under the Iuventus Plus grant (IP2014~025173)
and by Polish National Science Centre under the grants 2014/15/B/ST2/02528 and 2015/17/D/ST2/03530.
\end{acknowledgement}

\bibliography{refs}

\end{document}